\documentclass[conference]{IEEEtran}
\IEEEoverridecommandlockouts
\usepackage{cite}
\usepackage{comment}
\usepackage{amsmath,amssymb,amsfonts}
\usepackage{algorithmic}
\usepackage{graphicx}
\usepackage{textcomp}
\usepackage{svg}
\usepackage{subcaption}
\usepackage{adjustbox}
\usepackage{xcolor}
\usepackage{multirow}
\usepackage{booktabs}
\usepackage[section]{placeins}

\def\BibTeX{{\rm B\kern-.05em{\sc i\kern-.025em b}\kern-.08em
    T\kern-.1667em\lower.7ex\hbox{E}\kern-.125emX}}

\renewcommand{\baselinestretch}{1.0} 

\usepackage{flushend}
\newcommand{\myfigWidth}{0.7\columnwidth}

\begin{document}

\title{
 Classification of ECG based on Hybrid Features using CNNs for Wearable Applications

\thanks{*This work was supported in part by the China Scholarship Council, the Microelectronic Circuits Centre Ireland, and the Irish Research Council.}
}

\author{
    Li~Xiaolin$^{1}$,~\IEEEmembership{Student Member,~IEEE}, 
    Fang~Xiang$^{1}$,
    Rajesh~C.~Panicker$^{2}$~\IEEEmembership{Member,~IEEE},
    \\Barry Cardiff$^{1}$~\IEEEmembership{Senior Member,~IEEE},
    and Deepu John$^{1}$~\IEEEmembership{Senior Member,~IEEE}
    \\
    \\Emails: {\{xiaolin.li, fang.xiang\}}@ucdconnect.ie, rajesh@nus.edu.sg, {\{barry.cardiff, deepu.john\}}@ucd.ie
    
\thanks{$^{1}$University College Dublin,$^{2}$National University of Singapore.}%
}
\maketitle

\begin{abstract}
Sudden cardiac death and arrhythmia account for a large percentage of all deaths worldwide. 
Electrocardiography (ECG) is the most widely used screening tool for cardiovascular diseases. Traditionally, ECG signals are classified manually, requiring experience and great skill, while being time-consuming and prone to error. Thus machine learning algorithms have been widely adopted because of their ability to perform complex data analysis. Features derived from the points of interest in ECG - mainly Q, R, and S, are widely used for arrhythmia detection. In this work, we demonstrate improved performance for ECG classification using hybrid features and three different models, building on a 1-D convolutional neural network (CNN) model that we had proposed in the past. An RR interval features based model proposed in this work achieved an accuracy of 98.98\%, which is an improvement over the baseline model. To make the model immune to noise, we updated the model using frequency features and achieved good sustained performance in presence of noise with a slightly lower accuracy of 98.69\%. Further, another model combining the frequency features and the RR interval features was developed, which achieved a high accuracy of 99\% with good sustained performance in noisy environments. Due to its high accuracy and noise immunity, the proposed model which combines multiple hybrid features, is well suited for ambulatory wearable sensing applications.
\end{abstract}


\section{Introduction}

Traditionally, ECG recording is carried out using large non-portable equipment in professional medical institutions. With the development of low-cost wearable devices, it is feasible to acquire long-term ECG data and monitor user’s health in a real-time and cost-effective manner~\cite{nusJSENS_2019}. Abnormalities in ECG can be indicative of underlying cardiac issues, which can be detected effectively using signal processing and machine learning techniques~\cite{maryam_tbiocas2021}. Hence, there is a lot of interest in improving these techniques for a more reliable and robust cardiac monitoring using wearable sensors. The irregularity in cardiac rhythm, popularly known as arrhythmia, is one of the most popular conditions that can be detected in ECG and is very helpful in diagnosing and predicting cardiac anomalies. This work aims to improve the accuracy of ECG classification and minimize the effect of noise in the arrhythmia detection in ambulatory enviornments.

\begin{figure}[]
    \centering
    \includegraphics[width=\myfigWidth]{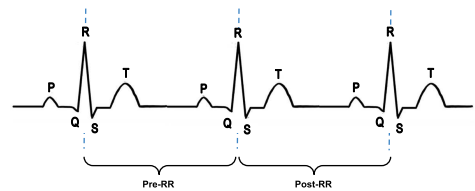}
    \caption{Illustrations of RR interval features extraction}
    \label{fig:RR_interval}
\end{figure}

Hannun~\emph{et~al.} proposed a huge deep neural network fed with raw single-lead ECG data to detect twelve rhythm classes using a very large dataset~\cite{andrew_arrhythmia}. This work used raw ECG data to detect arrhythmia, which while eliminates the need for feature detection and hence higher degree of automation, is not suited for resource constrained environments like wearable devices. RR interval (illustrated in Fig.~\ref{fig:RR_interval}), the time elapsed between two successive R-waves of the QRS signal on the electrocardiogram is widely recognized as a vital feature of ECG signals~\cite{RR_interval,9QRS_nus_2020}. Rahul~\emph{et~al.} utilized different classifiers to distinguish between normal, premature ventricular contraction (PVC), and premature atrial contraction (PAC) based on RR interval features, and achieved good performance~\cite{rrintervalbased}. While the model is simple due to the use of a single feature, it can only categorize the ECG data into three classes.
In a previous work, we extracted single lead ECG QRS complex as input to a convolutional neural network (CNN) as shown in Fig.~\ref{fig:CNN}. However, classes with fewer number of QRS complex has lower accuracy and sensitivity~\cite{xiaolin20201d}.
In this work, we add pre-RR interval, post-RR interval, average RR interval, and local mean RR interval to the model to improve the performance. In practice, ECG signal extraction is often affected by noise~\cite{noise_source_1,IoTJ21_Arlene}. We compute the frequency domain representation of the QRS complex using FFT, and concatenate it along with the signal as the input to the network to improve noise immunity. Finally, a method which combines the above two is also proposed. We introduce our baseline model in Section~\ref{Architecture}. Section~\ref{methods} introduces hybrid features methods, RR interval features and spectrum features are used to train the model. The results are detailed in Section~\ref{discussion} and Section~\ref{conclusion} concludes our work.

\section{\label{Architecture}Baseline Model Architecture}
\begin{figure}
    \centering
    \begin{subfigure}{\columnwidth}
    \includegraphics[width=\linewidth]{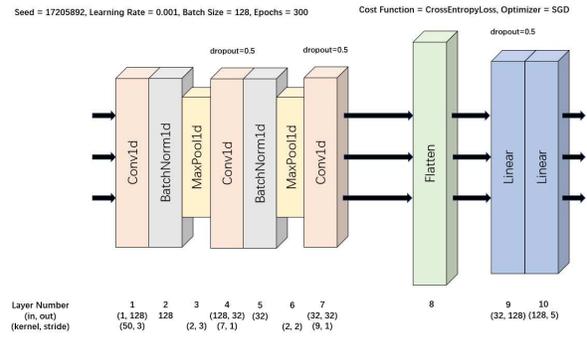}
    \caption{Baseline CNN architecture.}
    \label{fig:CNN}
    \end{subfigure}
    \centering
    \begin{subfigure}{\columnwidth}
    \includegraphics[width=\columnwidth]{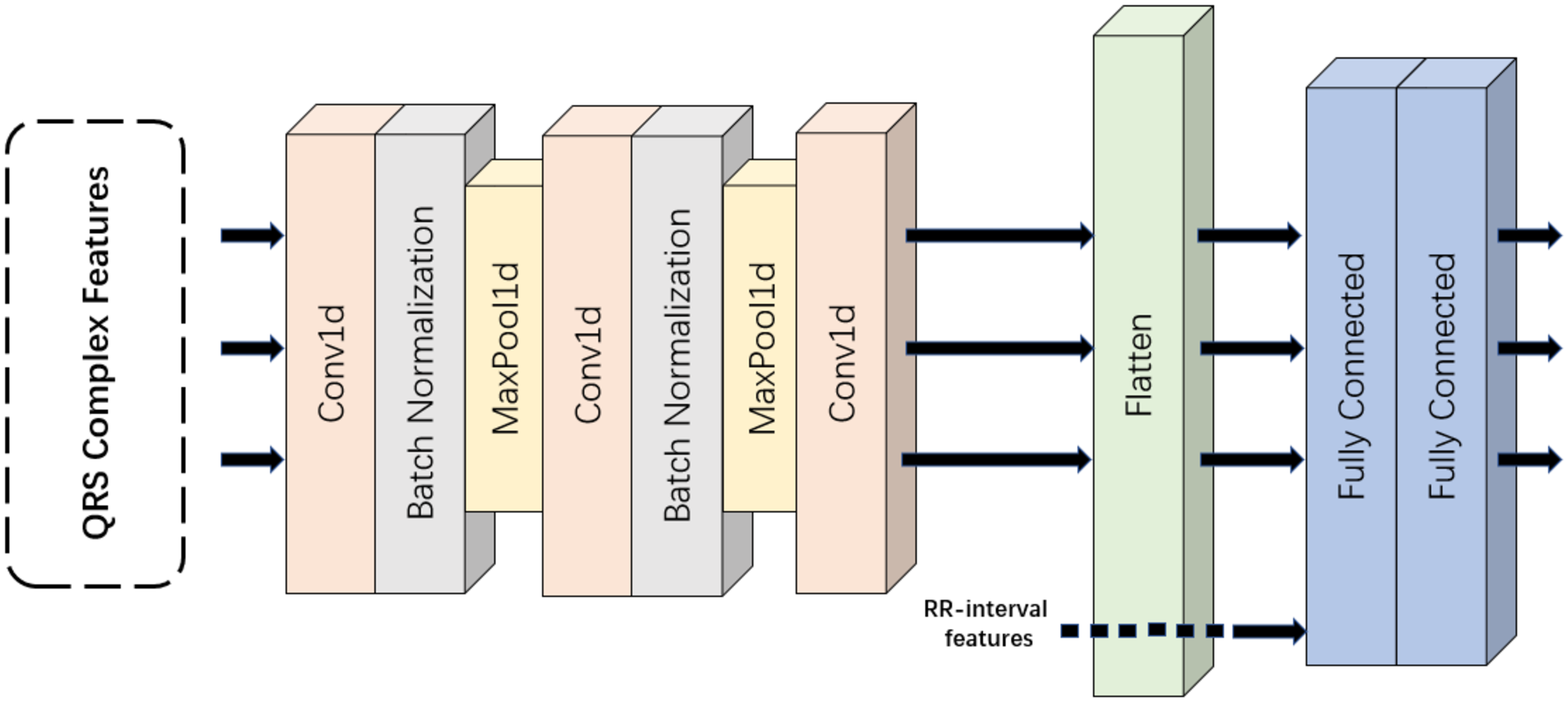}
    \caption{CNNe architecture.}
    \label{fig:CNNe}
    \end{subfigure}
    \centering
    \begin{subfigure}{\columnwidth}
    \includegraphics[width=\columnwidth]{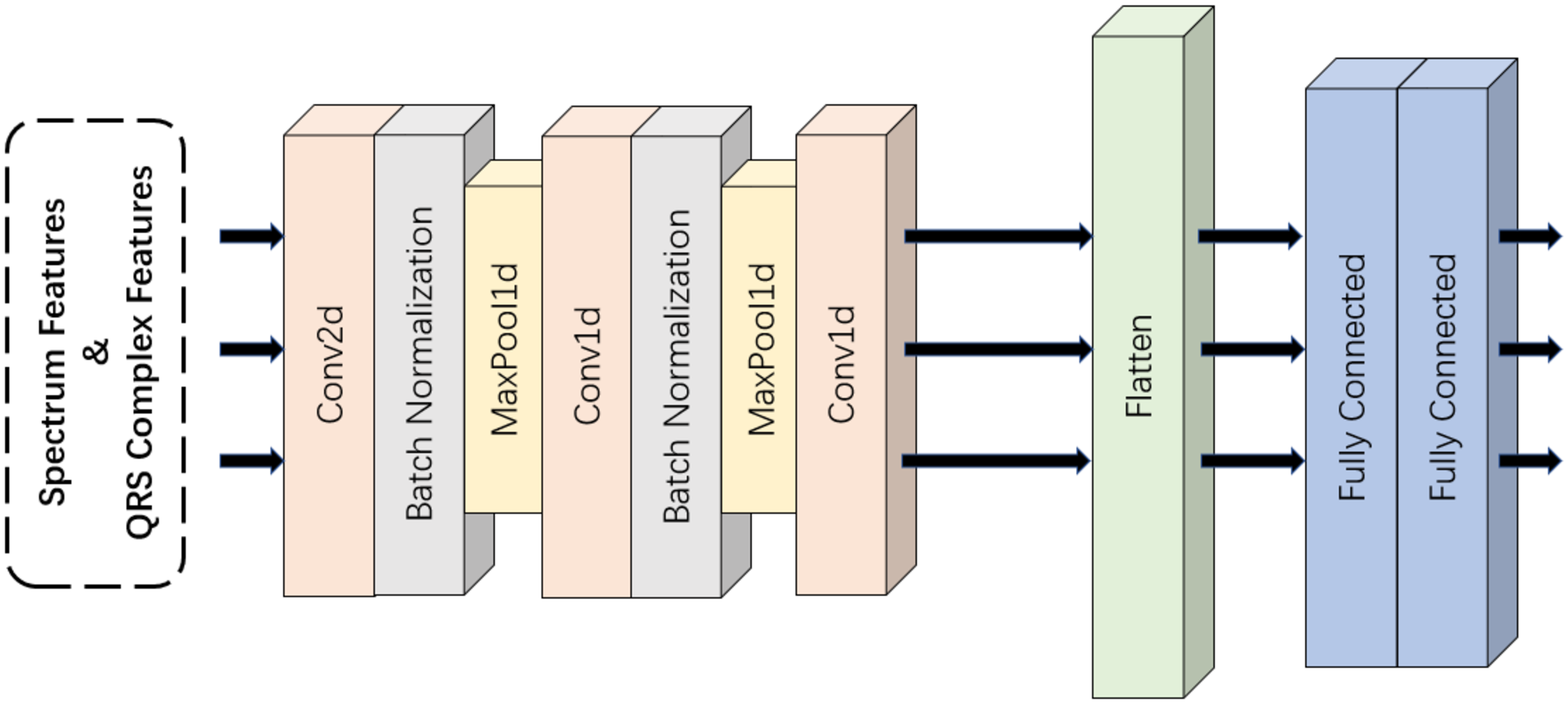}
    \caption{CNNf architecture.}
    \label{fig:CNNf}
    \end{subfigure}
    \centering
    \begin{subfigure}{\columnwidth}
    \includegraphics[width=\columnwidth]{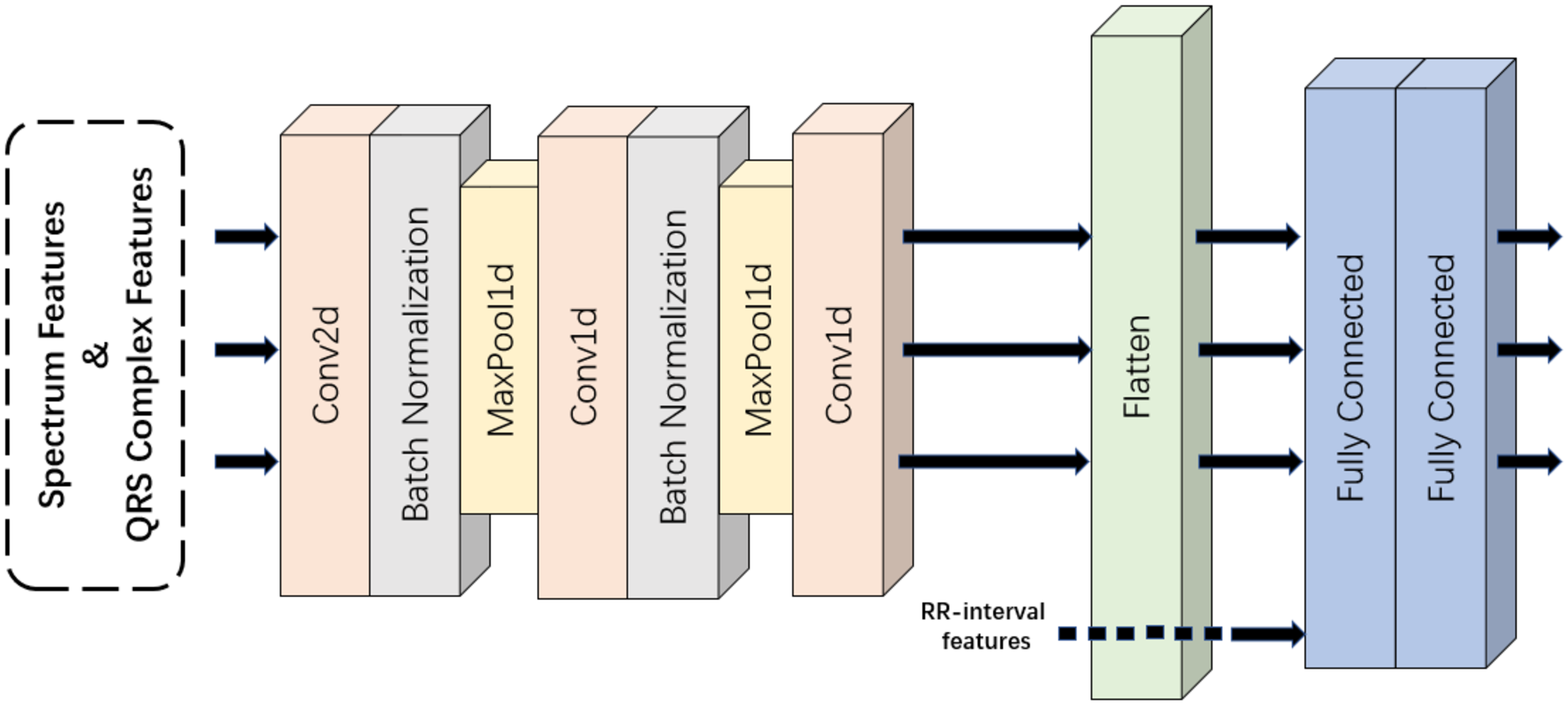}
    \caption{CNNef architecture.}
    \label{fig:CNNef}
    \end{subfigure}
    
    \caption{Baseline CNN architecture and the three proposed architectures.}
    \label{fig:architecture}
\end{figure}

Fig.~\ref{fig:CNN} illustrates a one dimensional CNN for heartbeat classification from single lead ECG~\cite{xiaolin20201d}. We use this CNN as our baseline model architecture. We use MIT-BIH Arrhythmia database~\cite{mitbih} and single MLII lead in this work. The input of CNN is a single QRS complex with 260 samples~\cite{260sample}, with half the samples taken before the R peak and the other half of the samples taken after the R peak, shown in Fig.~\ref{fig:CNN}. Each sample has 1/360 seconds, and the length of each input heartbeat is around 0.72 seconds. The output will be one of five specific heartbeat types based on AAMI standard~\cite{AAMI}. The model can classify five different types - N, SVEB, VEB, F, and Q - through these 10 layers. The original dataset is split into 70\%, 15\%, 15\% for training, validation, and testing purposes respectively. The ECG classification dataset obtained is unbalanced. Synthetic minority oversampling technique (SMOTE)~\cite{chawla2002smote} is used to balance the dataset by augmenting it with synthetic data.


\section{\label{methods}Hybrid Features based Architectures}
Amplitude and morphology of the QRS complex are important, but not the only characteristic for accurately classifying ECG signals. RR interval is also recognised as an important feature to detect anomalous ECG~\cite{ANNet,9QRS_nus_2020}. The ambulatory nature of ECG acquired from wearables, makes the amplitude and morphology highly susceptible to various kinds of noise. These include electrical activity of other muscles, shift in baseline caused by respiration, poor electrode contact, and interference from other electronic devices or equipment~\cite{noise_source_1, noise_source_2,noise_source_3}. However, RR intervals can be detected with acceptable accuracy even in the presence of severe noise~\cite{9QRS_nus_2020,IoTJ21_Arlene}. Hence, using RR interval features could enable a more accurate and robust classification of ECG data.

\subsection{RR interval \& QRS complex features}

The first architecture proposed, CNN-expanded (CNNe), is a variant of the baseline model which takes four additional RR interval features into the model compared to the baseline. The four RR intervals are described as follows. The pre RR interval is the length of the RR interval between the current QRS signal and the previous one. The post RR interval is the length of RR interval between the current QRS signal and the next one. The average RR interval is the mean value of all RR intervals in the record. Samples coming from the same record share the same value. The local mean RR interval is the mean value of ten successive R-waves of the QRS complex. These four features are appended into the first fully connected layer of the network based on our baseline model, shown in Fig.~\ref{fig:CNNe}.

\subsection{Spectrum \& QRS complex features}
\begin{figure}[!htb]
  \centering
  \includegraphics[width=\columnwidth]{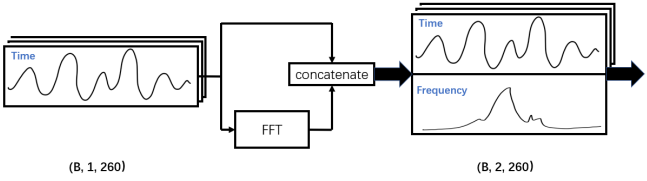}
  \caption{Spectrum features acquired method of input for CNNf architecture}
  \label{fig:CNNf_input}
\end{figure}

QRS complex morphology features are time series data and vulnerable to noise. Frequency features are less likely to be affected by various noise types, so we proposed CNN-fft (CNNf) architecture to utilize spectrum features of ECG. Spectrum sequences obtained from Fast Fourier Transform (FFT) are concatenated with the original time sequences to form the two-dimensional data input to the CNNf. FFT algorithm exploits symmetries for efficient calculation of discrete Fourier Transform (DFT). The transformation is most efficient when $n$ is a power of two. When the radix of FFT is not a power of two, we can use Bluestein’s algorithm~\cite{bluestein}. Fig.~\ref{fig:CNNf_input} shows the process of mixing two sequences of data using FFT. The spectrum and QRS complex features are combined to a 2-D sequence. Hence, the first layer of CNNf will be a 2-D convolutional layer, which differs from the baseline model as shown in Fig.~\ref{fig:CNNf}. In order to examine the robustness of these proposed architectures in the presence of noise, we introduced additional Gaussian White Noise (GWN) to the test dataset.

\subsection{RR interval \& Spectrum \& QRS complex features}

The third architecture shown in Fig.~\ref{fig:CNNef} is a hybrid of CNNe and CNNf, with the intention of having a model that can provide a good accuracy through the use of additional RR interval and spectrum features while being resistant to noise.

The spectrum features obtained from FFT is concatenated with the QRS time series, and forms the input to the network. The four RR interval features are appended to the result of the convolutional layers and forms the input to the first fully connected layer.

\section{\label{discussion}Results and Discussion}
The simulations were done using PyTorch on a GPU-accelerated computer. Table~\ref{tbl:params} displays parameters setting for each architecture in our work.

\begin{table*}[!htb]
\scriptsize
\caption{Parameters Setting of CNN, CNNe, CNNf, and CNNef.}
\centering
\scalebox{0.8}{
\begin{tabular}{ccccccccccc}
\hline
\textbf{\begin{tabular}[c]{@{}c@{}}(Input, Output)\\ (Kernel, Stride)\end{tabular}} & \textbf{Conv1d}                                                      & \textbf{Batch Normalization} & \textbf{Max Pooling} & \textbf{Conv1d}                                            & \textbf{Batch Normalization} & \textbf{Max Pooling} & \textbf{Conv1d}                                           & \textbf{Flatten} & \textbf{Fully Connected} & \textbf{Fully Connected} \\ \hline
CNN                                                                                 & \begin{tabular}[c]{@{}c@{}}(1, 128)\\ (50, 3)\end{tabular}           & (128)                        & (2, 3)               & \begin{tabular}[c]{@{}c@{}}(128, 32)\\ (7, 1)\end{tabular} & (32)                         & (2, 2)               & \begin{tabular}[c]{@{}c@{}}(32, 32)\\ (9, 1)\end{tabular} & ---              & (32, 128)                & (128, 5)                 \\
CNNf                                                                                & \begin{tabular}[c]{@{}c@{}}(1, 128)\\ ((50, 2), (3, 1))\end{tabular} & (128)                        & (2, 3)               & \begin{tabular}[c]{@{}c@{}}(128, 32)\\ (7, 1)\end{tabular} & (32)                         & (2, 2)               & \begin{tabular}[c]{@{}c@{}}(32, 32)\\ (9, 1)\end{tabular} & ---              & (32, 128)                & (128, 5)                 \\
CNNe                                                                                & \begin{tabular}[c]{@{}c@{}}(1, 128)\\ (50, 3)\end{tabular}           & (128)                        & (2, 3)               & \begin{tabular}[c]{@{}c@{}}(128, 32)\\ (7, 1)\end{tabular} & (32)                         & (2, 2)               & \begin{tabular}[c]{@{}c@{}}(32, 32)\\ (9, 1)\end{tabular} & ---              & (36, 128)                & (128, 5)                 \\
CNNef                                                                               & \begin{tabular}[c]{@{}c@{}}(1, 128)\\ ((50, 2), (3, 1))\end{tabular} & (128)                        & (2, 3)               & \begin{tabular}[c]{@{}c@{}}(128, 32)\\ (7, 1)\end{tabular} & (32)                         & (2, 2)               & \begin{tabular}[c]{@{}c@{}}(32, 32)\\ (9, 1)\end{tabular} & ---              & (36, 128)                & (128, 5)                 \\ \hline
\end{tabular}
}
\label{tbl:params}
\end{table*}

To evaluate the model, the dataset is split into training, validation and testing data as described in Section \ref{Architecture}. For each architecture (and the baseline model), we measure the overall accuracy, F1 score, sensitivity, specificity and precision. These are calculated using Equations~(\ref{eqn:Accuracy}), (\ref{eqn:Sen}), (\ref{eqn:Spe}) (\ref{eqn:Pre}), (\ref{eqn:f1}) separately, and displayed in Table~\ref{tbl:comparison4models}.

\begin{align}
    \text{Accuracy} &= \frac{TN+TP}{TN+TP+FP+FN} \label{eqn:Accuracy} \\
    \text{Sensitivity} &= \frac{TP}{TP+FN} \label{eqn:Sen}\\
    \text{Specificity} &= \frac{TN}{TN+FP} \label{eqn:Spe} \\
    \text{Precision} &= \frac{TP}{TP+FP} \label{eqn:Pre} \\
    \text{F1 Score} &= \frac{2\times \text{Sensitivity}\times \text{Precision}}{\text{Sensitivity} + \text{Precision}} \label{eqn:f1}
\end{align}
where,
\begin{itemize}
    \item TN = True Negative, the number of normal beats correctly classified as being normal.
    \item FN = False Negative, the number of non-normal(i.e. S, V, F, and Q) beats falsely classified as normal.
    \item TP = True Positive, the number of non-normal beats correctly classified.
    \item FP = False Positive, the number of non-normal beats incorrectly classified and normal beats falsely classified as non-normal.
\end{itemize}

\subsection{Without Noise}

Table~\ref{tbl:comparison4models} includes the results for CNN, CNNe, CNNf and CNNef for different performance metrics. 
The accuracy and F1 score of CNNe is slightly higher than that of the baseline CNN, while the sensitivity of CNNe is greater than that of CNN by 1.26\%. The results show that by adding RR interval features the performance can be improved. Since sensitivity is a very important figure of merit for biomedical applications, this improvement is generally beneficial. CNNef has a higher accuracy and F1 score than CNN as it includes RR interval features. Although its sensitivity is lower than that of CNNe, it is still higher than the baseline model.

\begin{table}[]
\scriptsize
\caption{Metrics of CNN, CNNe, CNNf and CNNef.}
\centering
\begin{tabular}{@{}cccccc@{}}
\toprule
      & Accuracy & F1 Score & Sensitivity & Specificity & Precision \\ \midrule
CNN   & 98.95\%  & 96.13\%  & 96.13\%     & 99.39\%     & 96.13\%   \\
CNNe  & 98.98\%  & 96.31\%  & 97.39\%     & 99.24\%     & 95.25\%   \\
CNNf  & 98.69\%  & 95.22\%  & 95.8\%      & 99.15\%     & 94.65\%   \\
CNNef & 99.00\%     & 96.32\%  & 96.36\%     & 99.41\%     & 96.27\%   \\ \bottomrule
\end{tabular}
\label{tbl:comparison4models}
\end{table}

\subsection{With Gaussian White Noise}
To evaluate noise immunity of each model, we added Guassian noise to the signal at different standard deviations ($\eta$), calculated using Equation~(\ref{eqn:SD}), with $\eta$ ranging from 1\% to 10\%. 

\begin{equation}
    Standard~Deviation = Average~Maximum\times \eta \times 0.3
\label{eqn:SD}
\end{equation}
where,
\begin{itemize}
    \item $Average~Maximum = 2.5mV$,
    \item $\eta = 1\%,2\%,...,10\%$.
\end{itemize}

The average value of the maximum amplitudes of the MLII lead signal in the dataset is approximately 2.5$mV$. Based on that, we evaluate the F1 score and sensitivity with different standard deviation levels. With standard deviation increments of 0.3\% of the average maximum amplitude (0.75$\mu V$), ten sets of noise are applied to the test set separately. 

\begin{figure}[!htb]
    \centering
    \includegraphics[width=\myfigWidth]{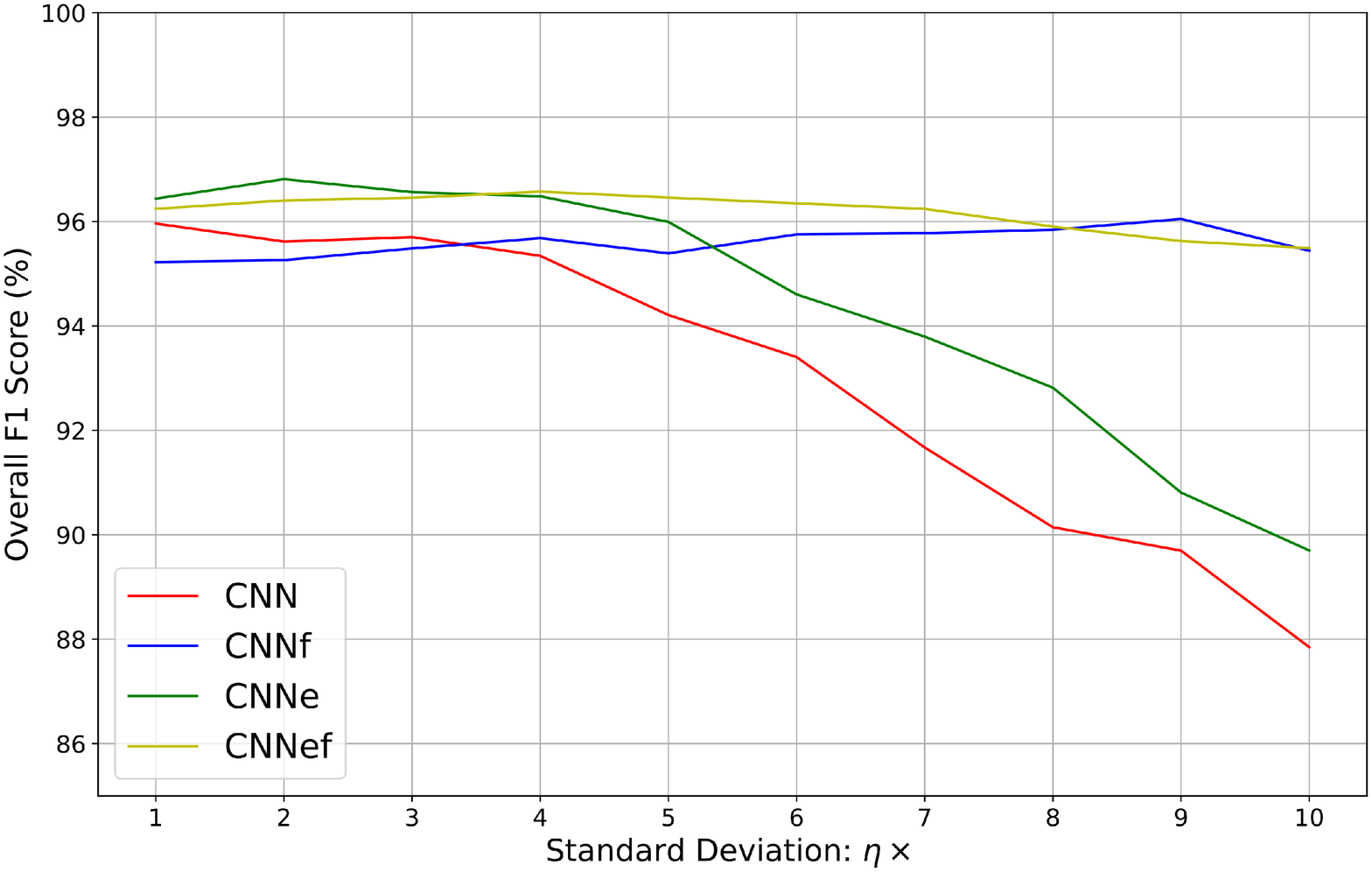}
    \caption{F1 score analysis with GWN.}
    \label{fig:f1_noise}
\end{figure}

\begin{figure}[!htb]
    \centering
    \includegraphics[width=\myfigWidth]{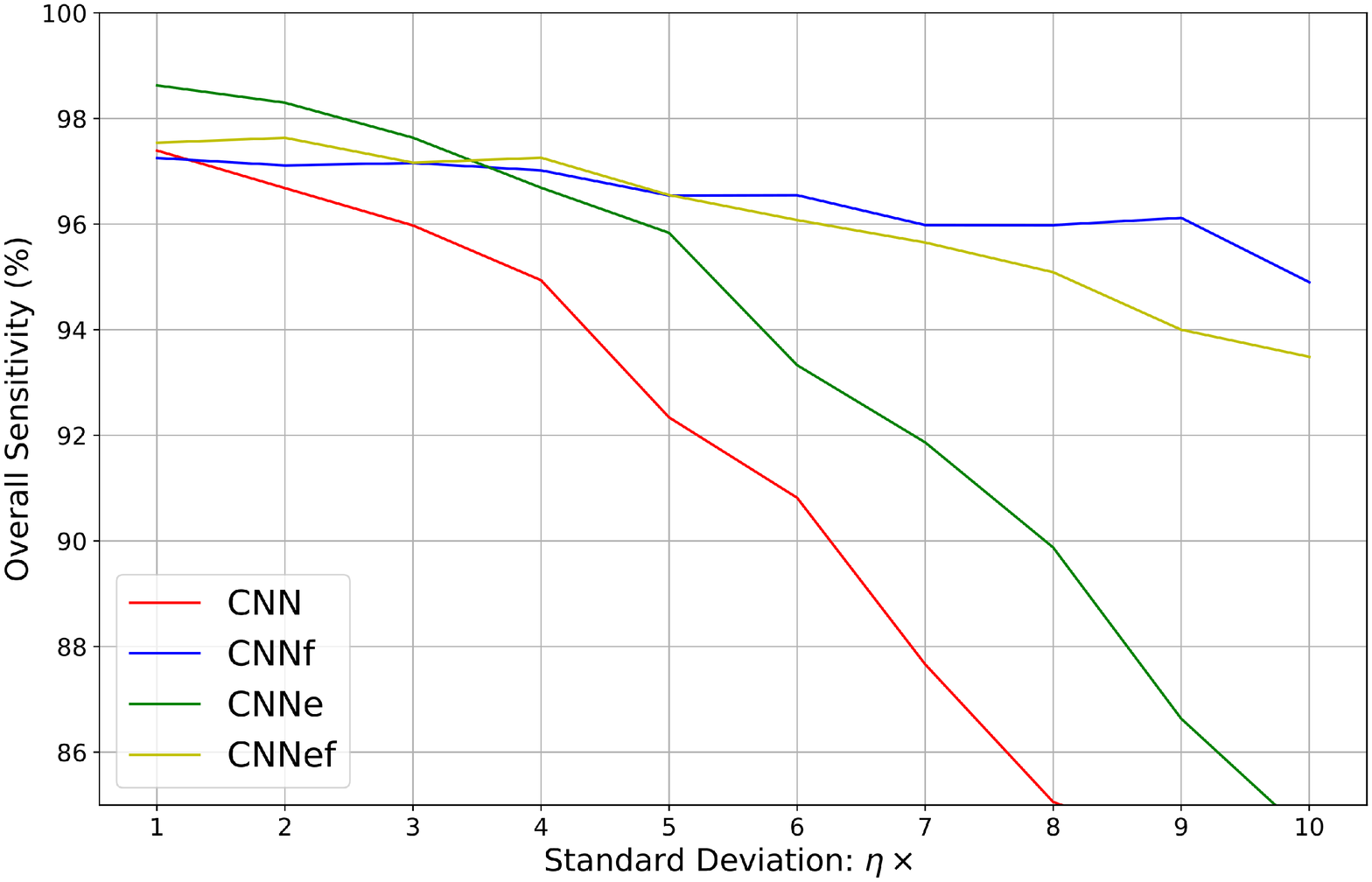}
    \caption{Sensitivity analysis with GWN.}
    \label{fig:sen_noise}
\end{figure}

Fig.~\ref{fig:f1_noise} and Fig.~\ref{fig:sen_noise} show the overall F1 score and sensitivity of CNN, CNNe, CNNf and CNNef when we add GWN with different standard deviations. It can be observed that the F1 score of CNN and CNNe drops sharply and is lower than the performance of CNNf when the standard deviation of noise increases. This indicates that the addition of frequency domain features indeed contribute to better performance in the presence of noise. However, the F1 score of CNNf and CNNef fluctuates in the range from 92\% to 94\% and does not fall as drastically as that of CNN and CNNe when the noise level is increased.

The sensitivity of CNN and CNNe drops substantially with noise increasing, even at relatively low noise levels. However, it can be observed that CNNf and CNNef suffers from a relatively lower drop. Although the sensitivity of CNNf decreases with the increase of noise standard deviation, it is still above 92\% in all the experiments.

\section{\label{conclusion}Conclusions and Future Work}
In this work, we proposed three different techniques - CNNe, CNNf and CNNef for ECG classification, and compared their performance with the baseline model CNN. The performance of these architectures was evaluated through the MIT-BIH Arrhythmia Database. CNNe adds four RR interval features to the fully connected layer and achieves better F1 score and sensitivity. CNNf concatenates spectral features with time domain data, which is shown to improve the noise resistance. CNNef is the combination of CNNe and CNNf which is shown to achieve higher performance as well as higher noise immunity than the baseline model. Future work includes a more extensive evaluation of the proposed CNNef strategy on other datasets as well as other architectures.
\ifCLASSOPTIONcaptionsoff
  \newpage
\fi

\bibliographystyle{IEEEtran}
\bibliography{bib_conf}

\end{document}